\documentclass[aps,prd,12pt,showpacs,notitlepage,nofootinbib,tightenlines]{revtex4-1}
\usepackage{amsmath}
\usepackage{bm}
\usepackage{times}
\usepackage{braket}
\usepackage{color}
\usepackage{epsfig}
\usepackage{slashed}
\usepackage{hyperref}
\usepackage{graphicx}
\usepackage{dcolumn}

\def\beq{\begin{equation}}
\def\eeq{\end{equation}}

\newcommand{\bee}{\begin{eqnarray}}
\newcommand{\eee}{\end{eqnarray}}
\newcommand{\non}{\nonumber\\ }
\newcommand\iden{\leavevmode\hbox{\small1\normalsize\kern-.33em1}}

\def \cpc{ {\bf Chin. Phys. C} }
\def \scp{{\bf Sci China Phys Mech $\&$ Astron} }

\def \epjc{{\bf Eur.Phys.J. C} }
\def \jpg{ {\bf J.Phys. G} }
\def \npb{ {\bf Nucl.Phys. B} }
\def \plb{ {\bf Phys.Lett. B} }
\def \pr{  {\bf Phys. Rept.} }
\def \prd{ {\bf Phys.Rev. D} }
\def \prl{ {\bf Phys.Rev.Lett.}  }
\def \epl{ {\bf EPL}  }

\def \zpc{ {\bf Z. Phys. C}  }
\def \jhep{ {\bf JHEP}  }

\begin{document}
\title{Single Higgs boson production at the ILC in the left-right twin Higgs model}
\author{Yao-Bei Liu$^{1,2}$, Zhen-Jun Xiao$^{1,3}$\footnote{Email: xiaozhenjun@njnu.edu.cn} }
\affiliation{1. Department of Physics and Institute of Theoretical Physics,
Nanjing Normal University, Nanjing 210023, P.R.China }
\affiliation{2. Henan Institute of Science and Technology, Xinxiang 453003, P.R.China}
\affiliation{3. Jiangsu Key Laboratory for Numerical Simulation of Large Scale
Complex Systems, Nanjing Normal University, Nanjing 210023, P.R. China}

\begin{abstract}
In this work, we analyze three dominant single SM-like Higgs boson production
processes in the left-right twin Higgs model (LRTHM): the Higgs-strahlung (HS) process
$e^{+}e^{-}\rightarrow Zh$, the vector boson fusion (VBF) process
$e^{+}e^{-}\to \nu\bar{\nu}h$ and the associate production with top
pair process $e^{+}e^{-}\to t\bar{t}h$ for three possible energy stages of the
International Linear Collider (ILC), and compared our results with the expected
experimental accuracies for various accessible Higgs decay channels.
The following observations have been obtained:
(i) In the reasonable parameter space, the LRTHM can generate moderate
contributions to theses processes with polarized beams;
(ii) Among various Higgs boson decay channels, the $b\bar{b}$ signal strength
is most sensitive to the LRTHM due to the high expected precision. For the
$t\bar{t}h$ production process, the absolute value of $\mu_{b\bar{b}}$ may
deviate from the SM prediction by over $8.7\%$ and thus may be detectable
at the future ILC with $\sqrt{s}=1$ TeV;
(iii) The future ILC experiments may give strong limit on the scale parameter $f$:
for the case of ILC-250 GeV, for example, the lower limit for parameter $f$ of the LRTHM
is $f > 1150$ GeV at the $2\sigma$ level.
\end{abstract}


\maketitle

\newpage
\section{Introduction}

On July 4th, 2012, a neutral Higgs boson with a mass around 125 GeV was discovered at
CERN's Large Hadron Collider (LHC) by both
the ATLAS and CMS collaborations \cite{atlas-1,cms-1}, whose properties appear to be
well consistent with those expected of
the Standard Model (SM) \cite{ichep1,ichep2}. However, the SM suffers from the so-called
little hierarchy problem \cite{0007265} and cannot provide a dark matter (DM) candidate,
which is actually a sound case for new physics (NP) beyond the SM. On the other hand,
the Higgs-like resonance with mass about 125 GeV can also
be well explained in many NP models where the Higgs is a pseudo-Goldstone boson.
Here we focus on the left-right twin Higgs model (LRTHM) \cite{ly,twin}, which can
successfully solve the little hierarchy problem and also predicts a good candidate
for weakly interacting massive particle (WIMP) dark matter.

As we know, the precision measurements of the Higgs boson are rather
challenging at the LHC, since various Higgs couplings to SM fermions
and vector bosons still have large uncertainties based on the current
LHC data \cite{lhc-1}. Thus the most precise measurements will be performed
in the clean environment of a future high energy $e^{+}e^{-}$ linear
collider, such as the International Linear Collider (ILC) \cite{ILC,1310.0763}.
Furthermore, a unique feature of the ILC is the presence of initial
state radiation (ISR) and beamstrahlung, which can help to improve
the measurement precision. The ILC is planned to operate at three
stages for the center-of-mass (c.m.) energy: 250 GeV, 500 GeV and 1
TeV. At the different energy stages, the Higgs-strahlung (HS) process
$e^{+}e^{-}\rightarrow Zh$, the vector boson fusion (VBF) process
$e^{+}e^{-}\rightarrow \nu\bar{\nu}h$, and the associated with top pair process
$e^{+}e^{-}\rightarrow t\bar{t}h$ are three main production channels for the
Higgs boson, which are very important for studying the properties of Higgs
boson and testing NP beyond the SM \cite{ilc-higgs}. These processes have
been studied in the context of the SM \cite{sm1,sm2} and various NP
models, such as the MSSM \cite{mssm}, the little Higgs models \cite{lht}
and other composite Higgs models \cite{other}.

The LRTHM predicted the existence of the new heavy gauge bosons,
top partner, neutral and charged scalars at or below the TeV scale, which can
produce rich phenomenology at the high energy colliders
\cite{Hock,0701071,dong,liu,ycx,liu1,wangl}. In the LRTHM, the couplings of the
electroweak gauge bosons to top quarks and the couplings of the SM-like Higgs
boson to ordinary particles are corrected at the order $\cal O$$(v^{2}/f^{2})$.
Besides, the new particles, such as the heavy neutral gauge boson and top partner,
can also contribute to some Higgs boson production processes. On the other hand,
the decays $h\rightarrow gg, \gamma\gamma$, and $Z\gamma$ all receive contributions
from the modified Higgs couplings and the new heavy particles, which has been
studied in our recent work \cite{liu-1311}. The aim of this paper is to consider
the processes $e^{+}e^{-}\rightarrow Zh$, $e^{+}e^{-}\rightarrow \nu\bar{\nu}h$
and $e^{+}e^{-}\rightarrow t\bar{t}h$
in the LRTHM, and see whether the effects of this model on these processes can
be detected in the future
ILC experiments.

The paper is organized as follows. In section II, we recapitulate the LRTHM
and lay out the couplings of the particles relevant to our calculation. In
Sec. III, we study the effects of the LRTHM on three single Higgs boson
production processes and project limits on the LRTHM from the future measurements
of the 125 GeV Higgs at the ILC with polarized beams. Finally, we give our conclusion
in Sec.IV.

\section{Relevant couplings in the LRTHM}

The twin Higgs mechanism was proposed as an interesting solution to the little
hierarchy problem, which can be implemented in
left-right model with the additional discrete symmetry being identified with
left-right symmetry \cite{ly}. Here we we will
briefly review the essential features of this model and focus on the couplings
relevant to our work. For more details of the LRTHM, one can see Ref.~\cite{Hock} and references therein.

The LRTHM has the gauged $SU(2)_{L}\times SU(2)_{R}\times U(1)_{B-L}$ sub-groups
of the global $U(4)\times U(4)$ symmetry. The left-right symmetry implies that
the gauge couplings of $SU(2)_{L}$ and $SU(2)_{R}$ are identical ($g_{2L}=g_{2R}=g$).
Two Higgs
fields ($H$ and $\hat{H}$) are introduced in the LRTHM, and each transforms as
$(4,1)$ and $(1,4)$, respectively under the global symmetry. They can be written as
\begin{eqnarray}
H=\left( \begin{array}{c} H_{L}\\ H_{R} \\
\end{array}  \right)\,,~~~~~~~~~~~~~~\hat{H}=\left( \begin{array}{c} \hat{H}_{L}\\
\hat{H}_{R} \\
\end{array}  \right)\,,
\end{eqnarray}
where $H_{L,R}$ and $\hat{H}_{L,R}$ are two component objects which
are charged under the $SU(2)_{L}\times SU(2)_{R}\times U(1)_{B-L}$
as
\begin{equation}
H_{L}~and~ \hat{H}_{L}: (2, 1, 1),~~~~~~~~H_{R}~ and~ \hat{H}_{R}:(1, 2, 1).
\end{equation}
 The global $U(4)_{1}(U(4)_{2})$
symmetry is spontaneously broken down to its subgroup
$U(3)_{1}(U(3)_{2})$ with non-zero vacuum expectation values(VEV):
\beq
<H>=(0,0,0,f)^{\rm T}, \quad  <\hat{H}>=(0,0,0,\hat{f})^{\rm T},
\eeq
The spontaneous symmetry breaking results in 14 Nambu-Goldstone bosons, which can be
parameterized as described in Ref. \cite{Hock}. The gauge symmetry
$SU(2)_{L}\times SU(2)_{R}\times U(1)_{B-L}$ is eventually broken down to $U(1)_{EM}$.
Six Goldstone bosons are eaten by the SM gauge bosons $(W^{\pm}, Z)$ and by the heavy
gauge bosons $(W^{\pm}_{H},Z_{H})$. Here $W_{H}^{\pm}$ and $Z_{H}$ are three additional
gauge bosons with masses of a few TeV in this model.
The remaining eight scalars include: one SM-like Higgs boson h, one neutral pseudoscalar
$\phi^{0}$, a pair of charged scalar $\phi^{\pm}$, and a $SU(2)_{L}$ doublet
$\hat{h}=(\hat{h}_{1}^{+},\hat{h}_{2}^{0})$. Noticed that a parity is introduced
in this model under which $\hat{H}$ is odd while all other fields are even,
which forbids renormalizable couplings between $\hat{H}$ and fermions.
The lightest particle in the odd $\hat{h}_{2}^{0}$ is stable and can be treated
as a candidate for dark matter, which has been studied in Ref.\cite{dm}.

Besides the new heavy gauge bosons, a pair
of vector-like quarks are also introduced to cancel the one-loop quadratic
divergence of Higgs mass.
The masses of the heavy gauge bosons, SM-like top quark and top partner are
given by \cite{Hock}
\begin{eqnarray}
m_{Z_{H}}^{2}&=& \frac{e^{2}C_{W}^{2}}{2S_{W}^{2}C_{2W}}(f^{2}+\hat{f}^{2})-M_{Z}^{2},
\label{eq:mzh01}\\
M_{W_{H}}^{2}&=& \frac{1}{2}g^{2}(\hat{f}^{2}+f^{2}\cos^{2}x),\\
m_{t}^{2}&=& \frac{1}{2}(M^{2}+y^{2}f^{2}-N_{t}), \\
m_{T}^{2}&=& \frac{1}{2}(M^{2}+y^{2}f^{2}+N_{t}),\label{eq:mt}
\end{eqnarray}
with $N_{t}=\sqrt{(M^{2}+y^{2}f^{2})^{2}-y^{4}f^{4}\sin^{2}2x}$ and
$x=v/(\sqrt{2}f)$. The values of the energy scales $f$ and $\hat{f}$ are interconnected once
we set $v\simeq246$ GeV. The parameter $M$ is essential to the mixing between the
SM-like top quark $t$ and its partner $T$. The value of $y$ can be determined
by fitting the experimental value of the
SM-like top quark mass $m_{t}$.
The abbreviations $S_{W}$, $C_{W}$ and $C_{2W}$ in Eq.~(\ref{eq:mzh01}) are
defined in the following way:
\begin{eqnarray}
S_{W}= \sin\theta_{W}= \frac{g'}{\sqrt{g^{2}+2g'^{2}}},  \quad
C_{W}= \cos\theta_{W}= \sqrt{\frac{g^{2}+g'^{2}}{g^{2}+2g'^{2}}},
\quad C_{2W}= \cos 2\theta_{W},
\label{eq:df01}
\end{eqnarray}
where $\theta_W$ is the well-known Weinberg angle.
The unit of the electric charge can then
be written in the form of $e=gS_{W}=gg'/\sqrt{g^{2}+2g'^{2}}$.

The relevant couplings for the vertices in the LRTHM used in this work are
given as follows  \cite{Hock}:
\bee
 g_{L}^{Zt\bar{T}}&=&\frac{eC_{L}S_{L}}{2S_{W}C_{W}}, \qquad
 g_{R}^{Zt\bar{T}}=\frac{ef^{2}x^{2}S_{W}C_{R}S_{R}}{2\hat{f}^{2}C^{3}_{W}},\\
g_{L}^{Z_{H}t\bar{T}}&=&\frac{eC_{L}S_{L}S_{W}}{2C_{W}\sqrt{C_{2W}}}, \qquad
g_{R}^{Z_{H}t\bar{T}}=-\frac{eC_{W}C_{R}S_{R}}{2S_{W}\sqrt{C_{2W}}},\\
 g_{L}^{Z_{H}e^{+}e^{-}}&=&\frac{2eS_{W}}{4C_{W}\sqrt{C_{2W}}}, ~~~~~~
 g_{R}^{Z_{H}e^{+}e^{-}}=\frac{e(1-3C_{2W})}{4S_{W}C_{W}\sqrt{C_{2W}}},\\
V_{\phi^{0}t\bar{t}}&=&-\frac{iy}{\sqrt{2}}S_{L}S_{R},~~~~~~~~~~
V_{h\phi^{0}Z_{\mu}}=\frac{iex}{6S_{W}C_{W}}p_{Z\mu},\\
V_{ht\bar{T}}&=&-\frac{y}{\sqrt{2}}[(C_{L}S_{R}+S_{L}C_{R}x)P_{L}
+(C_{L}S_{R}x+S_{L}C_{R})P_{R}],\\
V_{hZ_{H_{\mu}}Z_{H_{\nu}}}&=&-\frac{e^{2}fx}{\sqrt{2}S^{2}_{W}C^{2}_{W}}g_{\mu\nu},~~~~~~~
V_{hZ_{\mu}Z_{H_{\nu}}}=\frac{e^{2}fx}{\sqrt{2}C^{2}_{W}\sqrt{C_{2W}}}g_{\mu\nu},
 \eee
with
\begin{eqnarray}
S_{L}&=&\frac{1}{\sqrt{2}}\sqrt{1-(y^{2}f^{2}\cos 2x+M^{2})/N_{t}},\quad
C_{L}=\sqrt{1-S^{2}_{L}},\\
S_{R}&=&\frac{1}{\sqrt{2}}\sqrt{1-(y^{2}f^{2}\cos 2x-M^{2})/N_{t}},\quad
C_{R}=\sqrt{1-S^{2}_{R}}.
\end{eqnarray}
Here the momentum $p_{Z}$ in the coupling $V_{h\phi^{0}Z}$ in Eq.~(12) refers to the
incoming momentum of $Z$ boson.

In the LRTHM, the normalized couplings of $hf\bar{f} (f=b,c), ht\bar{t},
h\tau^{+}\tau^{-}, hVV^{\ast} (V=Z,W), hgg$, and $h\gamma\gamma$ are given
by \cite{Hock,wangl}:
\bee
V_{\rm hVV}/{\rm SM} &\equiv& \frac{V_{\rm HVV}}{V^{\rm SM}_{\rm HVV}}= 1-\frac{v^{2}}{6f^{2}},\\
V_{\rm hf\bar{f}}/{\rm SM}&=&\frac{V_{\rm h\tau^{+}\tau^{-}}}{V^{\rm SM}_{\rm h\tau^{+}\tau^{-}}}
=1-\frac{2v^{2}}{3f^{2}},\qquad
V_{\rm ht\bar{t}}/{\rm SM} =C_{L}C_{R},\\
V_{\rm hgg}/{\rm SM} &=& \frac{\frac{1}{2}F_{1/2}(\tau_{t})y_{t}+\frac{1}{2}F_{1/2}(\tau_{T})y_{T}}{\frac{1}{2}F_{1/2}(\tau_{t})}, \\
V_{\rm h\gamma\gamma}/{\rm SM} &=& \frac{\frac{4}{3}F_{1/2}(\tau_{t})y_{t}+\frac{4}{3}F_{1/2}(\tau_{T})y_{T}+F_{1}(\tau_{W})y_{W}}{\frac{4}{3}F_{1/2}(\tau_{t})+F_{1}(\tau_{W})},
\eee
with
\bee
F_{1}&=&2+3\tau+3\tau(2-\tau)f(\tau),\quad
F_{1/2}=-2\tau[1+(1-\tau)f(\tau)], \non
f(\tau)&=&\left [\sin^{-1}(1/\sqrt{\tau}) \right ]^{2},~~~~~~~~~~\quad
g(\tau)= \sqrt{\tau-1}\sin^{-1}(1/\sqrt{\tau}),
\eee
for $\tau_{i}=4m_{i}^{2}/m_h^2 \geq  1$. The relevant couplings $ y_{t}$ and $y_{T}$
can be written as
\beq
y_{t}= S_{L}S_{R},~~~~~~~~~~~~~~~~~~\quad y_{T}= \frac{m_{t}}{m_{T}}C_{L}C_{R},
\eeq
which can be determined by the parameters $f$ and $M$. Here we have neglected the
contributions from $W_H$ and $\phi^\pm$ for the $h\to  \gamma\gamma$ decay, this
is because their contributions are even much smaller than that for the T-quark
\cite{liu-1311}.
On the other hand, the relation between $G_{F}$ and $v$ is modified from its SM
form, introducing an
 additional correction $y_{G_{F}}$ as $1/v^{2}=\sqrt{2}G_{F}y^{2}_{G_{F}}$ with
 $y^{2}_{G_{F}}=1-v^{2}/(6f^2)$.

\section{Numerical results and discussions}

In the LRTHM, the tree-level Feynman diagrams of the processes
$e^{+}e^{-}\rightarrow Zh$, $e^{+}e^{-}\rightarrow \nu\bar{\nu}h$,
and $e^{+}e^{-}\rightarrow t\bar{t}h$ are shown in Figs. 1-2. We can
see that the modified couplings of $hXX$ and the additional particles
($Z_{H}$ and $T$-quark) can contribute to these processes at the tree level.

\begin{figure}[th]
\vspace{-2cm}
\begin{center}
\leftline{\epsfxsize=18cm\epsffile{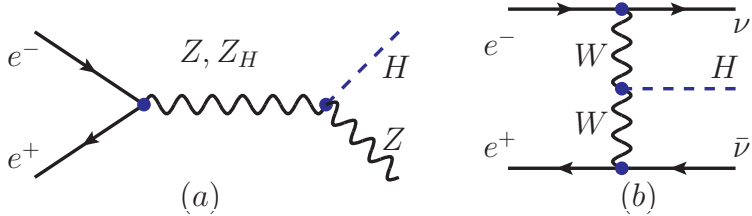} }
\end{center}
\vspace{-19cm}
\caption{\small Feynman diagrams for the Higgs-strahlung process
 $e^{+}e^{-}\rightarrow Zh$ (left) and $WW$-fusion process
 $e^{+}e^{-}\rightarrow \nu\bar{\nu}h$ (right).} \label{fig1}
\end{figure}

Obviously, the heavy $T$-quark, the neutral scalar $\phi^{0}$ and the
charged Higgs bosons $\phi^{\pm}$ can also contribute to these processes at the
loop level. However, their loop contributions are all most possibly very small and can be
neglected due to the following two reasons:
\begin{enumerate}
\item[ (i)]
The $T$-quark is very heavy and meanwhile the couplings of $hT\bar{T}$ is very
small as shown in Ref.~\cite{Hock}.

\item[(ii)]
The couplings of $h\phi^{0}\phi^{0}$ and $h\phi^{+}\phi^{-}$ are both suppressed largely by a factor
of $v^{2}/(2f^{2}) < 0.08$.
\end{enumerate}
We therefore only focus on those tree contributions at the lowest-order in this work.
We will evaluate the relevant loop contributions in the near future to examine the smallness
of the loop contributions directly.

\begin{figure}[th]
\vspace{-2cm}
\begin{center}
\leftline{\epsfxsize=14cm\epsffile{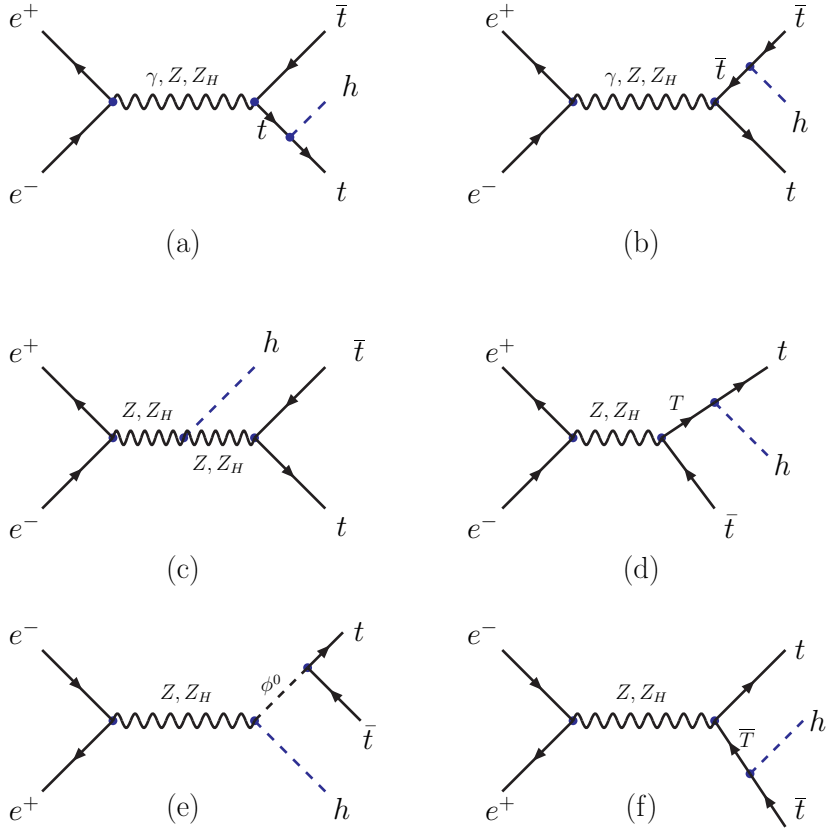} }
\end{center}
\vspace{-7cm}
\caption{\small Feynman diagrams for the process
$e^{+}e^{-}\rightarrow t\bar{t}h$ in the LRTHM.}
\end{figure}

In our numerical calculations, the SM-like Higgs boson mass is fixed as 125 GeV,
the SM input parameters involved are taken from \cite{data}.
Besides, there are two LRTHM parameters: $f$ and $M$.
The indirect constraints on $f$ come from the $Z$-pole precision measurements,
the low energy neutral current process and the high energy precision
measurements off the $Z$-pole: all these data prefer the parameter $f$ to be larger
than 500-600 GeV \cite{Hock}. The mixing parameter $M$ also be constrained by the $Z\to b\bar{b}$ branching ratio
and oblique parameters \cite{Hock,0701071}.

In the LRTHM, furthermore, the masses of heavy neutral gauge boson $Z'$ and top partner $T$
are determined by the given values of $f$ and $M$.
Currently, the masses of the new heavy particles, such as the $Z'$ and $T$
have been constrained by the LHC experiments, as described in
Refs.~\cite{bound-t,bound-zh}. In other words, the LHC data also imply
some indirect constraints on the allowed ranges of both the parameters $f$
and $M$ through their correlations with $m_{\rm Z'}$ and $m_{\rm T}$, as discussed in Ref.~\cite{liu-1311}.
For example, the top partner $T$ with mass below $656$ GeV are excluded at $95\%$
confidence level according to the ATLAS data \cite{atlas-3}
if one takes the assumption of a branching ratio $BR(T\rightarrow W^{+}b)=1$.
The ATLAS \cite{atlas-ztt} and CMS \cite{cms-ztt} collaborations have excluded the
leptophobic $Z'$ boson with the mass smaller than $1.32$ TeV (ATLAS) and $1.3$
TeV (CMS). A $Z'$ in the LRTHM with a mass below $940$ GeV has been excluded
in Ref.~\cite{lrth-z1} by using the D0 and CDF measurements.
By taking the above constraints from the electroweak precision measurements and the LHC data into account,
we here assume that the values of the parameter $f$ and $M$ are in the ranges of
\beq
600 \rm GeV \leq f \leq 1500 \rm GeV, \quad 0 \leq M \leq 150 \rm GeV,
\eeq
in our numerical evaluations.

From Fig.2(e) we can see that the neutral scalar $\phi^{0}$ can also contribute to
the cross section. However, this contribution is very small due to the suppressed
couplings of $h\phi^{0}Z$  and $\phi^{0}t\bar{t}$, and thus we can safely take
its mass as $m_{\phi^{0}}=150$ GeV.
All the numerical evaluations have been done by using the CalcHEP package \cite{calchep}.

\subsection{The production cross section with polarized beams}

As is well-known, beam polarization is an essential ingredient at future high-energy linear collider experiments.
Since the electrons and positrons in the beams are essentially chirality eigenstates, appropriate beam polarization
in some processes can greatly increase the new physics signals and reduce the SM background \cite{pol}.
With the longitudinal polarization of the initial electron and positron beams, the cross section of a process
can be expresses as \cite{0507011}
\bee
\sigma(P_{e^{-}},P_{e^{+}})&=& \frac{1}{4}[(1+P_{e^{-}})(1+P_{e^{+}})\sigma_{RR}+(1-P_{e^{-}})(1-P_{e^{+}})\sigma_{LL} \nonumber \\
 &&
 +(1+P_{e^{-}})(1-P_{e^{+}})\sigma_{RL}+(1-P_{e^{-}})(1+P_{e^{+}})\sigma_{LR}],
\eee
where $P_{e^{-}}(P_{e^{+}})$ is the polarization degree of the electron (positron) beam.
$\sigma_{LR}$ stands for the cross section for completely left-handed polarized $e^{-}$ beam ($P_{e^{-}}=-1$)
and completely right-handed polarized $e^{+}$ beam ($P_{e^{+}}=1$),
and other cross sections $\sigma_{LL}$, $\sigma_{RR}$, and $\sigma_{RL}$ are defined analogously.

\begin{figure}[thb]
\begin{center}
\vspace{-0.5cm}
\scalebox{1.2}{\epsfig{file=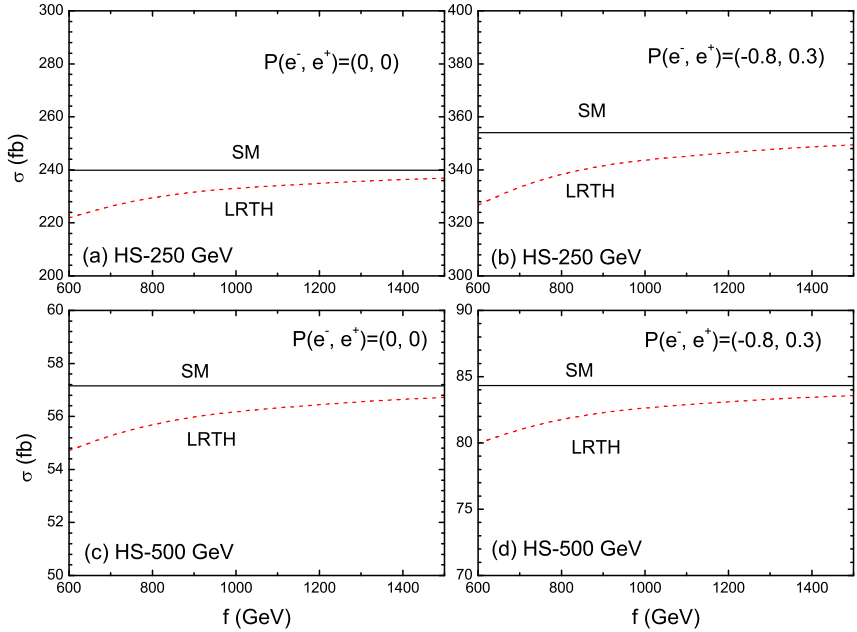}}
\vspace{-0.5cm}
\end{center}
\caption{\small The cross sections $\sigma$ versus the scale
$f$ at the ILC with unpolarized and polarized beams for HS process with $M=150$ GeV. }
\end{figure}

\begin{figure}[thb]
\begin{center}
\vspace{-0.5cm}
\scalebox{1.2}{\epsfig{file=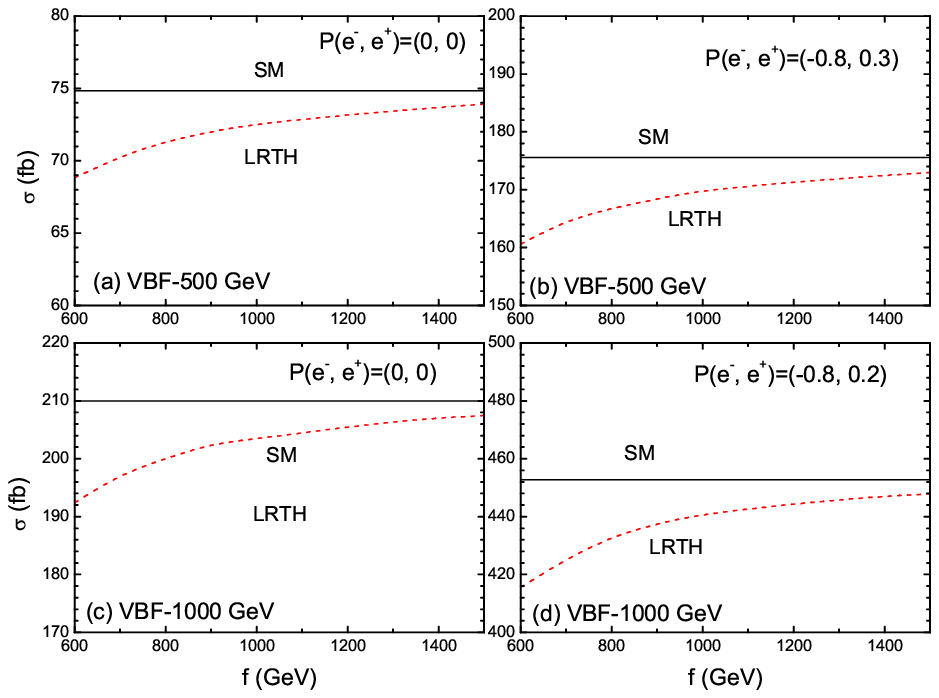}}
\vspace{-1cm}
\scalebox{1.2}{\epsfig{file=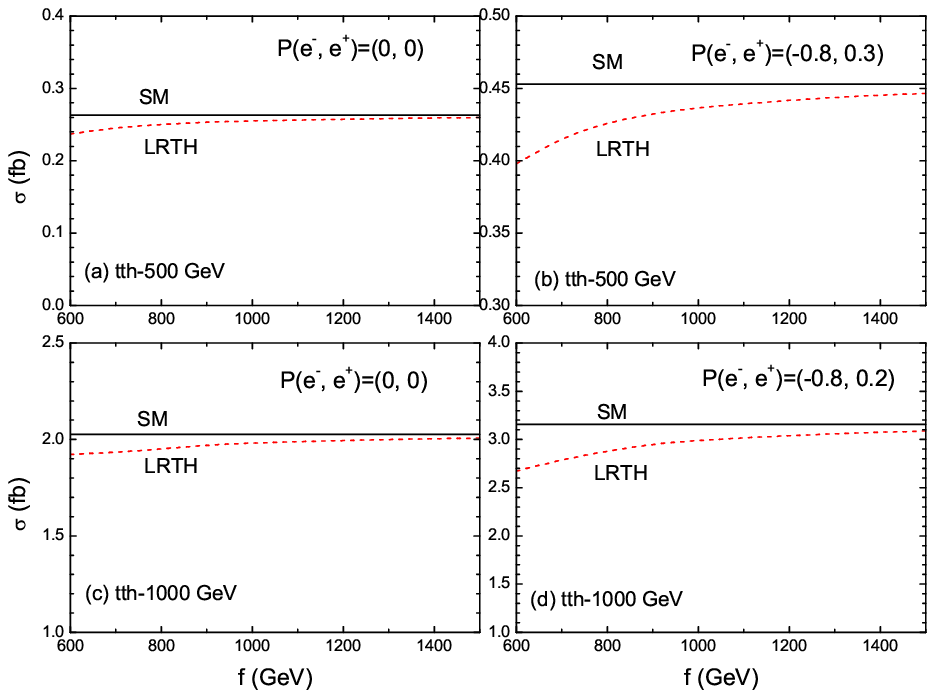}}
\end{center}
\caption{\small The same as Fig.~3, but for the case of VBF and $t\bar{t}h$ processes.}
\end{figure}

In Figs.~3 and 4, we plot the production cross sections for three processes $e^+e^-\to (Zh, \nu\bar{\nu}h,t\bar{t}h)$
with $\sqrt{s}=250,500$ GeV and $1000$ GeV at the ILC in the SM and LRTHM.
One can see that the cross sections in LRTHM are always smaller than those in the SM, and the values of
cross sections increase as the parameter $f$ increases,
which means that the correction of the LRTHM decouples with the scale $f$ increasing.
This is similar with the situation in the little Higgs models \cite{little1,little2}.
For comparison we also show the corresponding results for unpolarized beams.
One can see that the cross sections with polarized beams are always larger than those with the unpolarized beams,
and thus make the ILC more powerful in probing such new physics effects.

For HS process, the Feynman diagram involving $s$-channel gauge bosons exchange have more contributions
to $\sigma_{LR}$ than to $\sigma_{RL}$, this is because the neutral gauge bosons do not couple to
$e^{-}_{R}e^{+}_{R}$ and $e^{-}_{L}e^{+}_{L}$, while the couplings to $e^{-}_{L}e^{+}_{R}$
are stronger than those to $e^{-}_{R}e^{+}_{L}$.
For the case of the unpolarized beams with $\sqrt{s}=250 (500)$ GeV, the cross section of the HS process is about 240 (57) $fb$,
which is similar with the results in Refs.~\cite{1309.4819}.
At low energy (such as $\sqrt{s}=250$ GeV), the HS process is dominant and the cross section can reach about
350 fb for the case of the polarized beams, while the VBF and $t\bar{t}h$ processes are more significant
at higher energies.

\begin{figure}[ht]
\begin{center}
\scalebox{0.70}{\epsfig{file=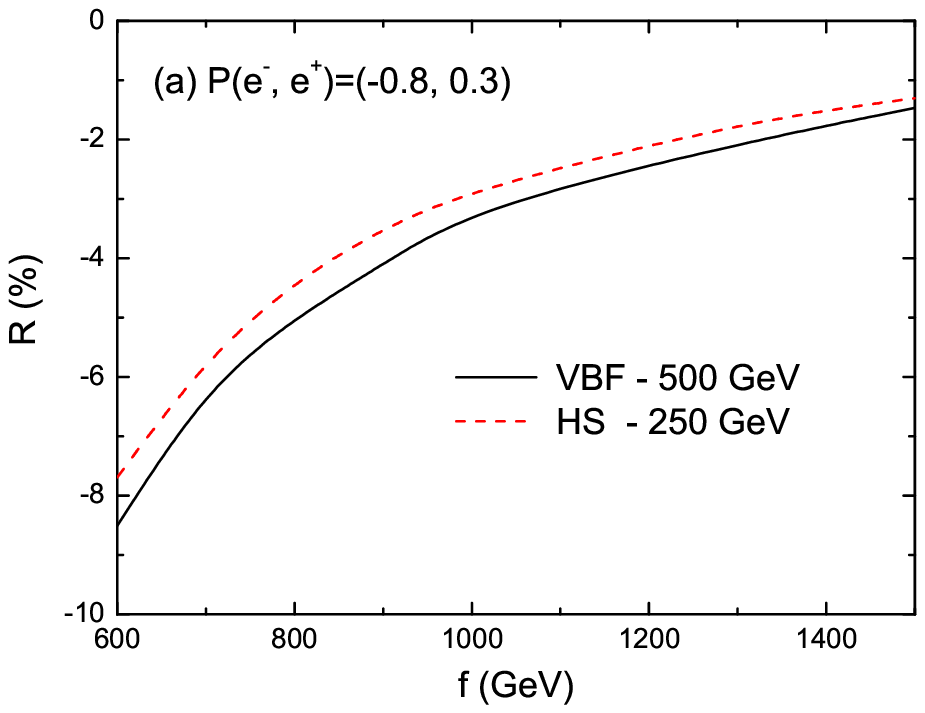}\epsfig{file=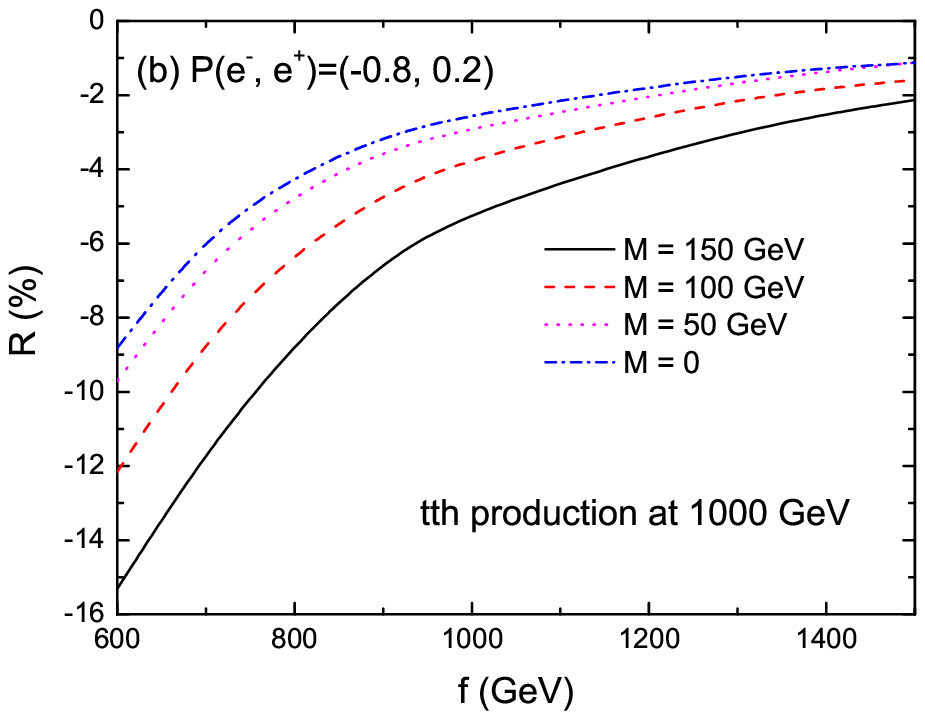}}
\end{center}
\caption{\small The relative correction parameter
$R=(\sigma^{\rm LRTHM}-\sigma^{\rm SM})/\sigma^{\rm SM}$ versus the scale
$f$ at the ILC with polarized beams for three processes. }
\end{figure}

In Fig.~5, we show the $f-$dependence of the relative correction parameter
$R=(\sigma^{\rm LRTHM}-\sigma^{\rm SM})/\sigma^{\rm SM}$
for three production channels with polarized beams. For the considered
three processes, the relative corrections are all negative and has
a moderate dependence on the variations of the scale parameter $f$.
The absolute values of suppression are larger
than $5\%$ for small values of $f$, but become smaller  for larger values of the scale $f$.
The total SM electroweak correction for the HS production process is
about $5\%$ for $m_{h}=125$ GeV and $\sqrt{s}=250$ GeV \cite{sm1}.
The expected accuracies for HS and VBF processes are about $2.0\sim2.6\%$
for $m_{h}=125$ GeV \cite{1310.0763}.
For the process $e^{+}e^{-}\to t\bar{t}h$, the relative corrections are
sensitive to the variations of the parameters $M$ and $f$, and become larger in size
for larger $M$ and smaller $f$.
For $M=150$ GeV and $f\leq 800$ GeV, for example,
the NP contribution can alter the SM cross section over $8\%$.
At the ILC with $\sqrt{s}=1000$ GeV, an accuracy of about $6.3\%$ could  be
reached with the polarized beams \cite{1310.0763}.
The new physics effects considered here therefore might be detected
in the future high precision ILC experiments.

\subsection{The Higgs signal strengths}

 Considering the Higgs boson decay channels, the Higgs signal strengths can be defined as
\beq
\mu_{i}=\frac{\sigma_{\rm LRTHM}\times BR(h\to i)_{\rm LRTHM}}{\sigma_{\rm SM}
\times BR(h\to i)_{\rm SM}},
\eeq
where $i$ denotes a possible final state of the Higgs boson decay (for
example $b\bar{b}$, $gg$, $c\bar{c}$, $ZZ^{\ast}$ and $\gamma\gamma$).
The projected $1\sigma$ sensitivities of the relevant channels at the ILC are shown
in Table I. We can see that the $b\bar{b}$ channel is more easily accessible
than other channels.

\begin{table}[ht]
\begin{center}
\caption{ Projected $1\sigma$ sensitivities of various channels for the ILC
operating at $\sqrt{s} = 250$ GeV, 500 GeV and 1000 GeV, respectively \cite{1310.0763,1306.6352}. }
\label{table1} \vspace{0.2cm}
\begin{tabular}{|c|c|c|c|c|c|c|c|}
\hline $\sqrt{s}$ &\multicolumn{2}{c|}{ 250 GeV}  &\multicolumn{3}{c|}{500 GeV }&\multicolumn{2}{c|}{1 TeV }\\
$(P_{e^{-}}, P_{e^{+}})$ &\multicolumn{2}{c|}{  (-0.8, 0.3) }&\multicolumn{3}{c|}{ (-0.8, 0.3)}&\multicolumn{2}{c|}{ (-0.8, 0.2)}\\
\hline
channel& HS &VBF& HS & VBF & ttH & VBF & ttH\\
\hline
$h\rightarrow b\bar{b}$ & $1.1\%$ &$10.5\%$ & $1.8\%$ & $0.66\%$ & $35\%$ & $0.47\%$ & $8.7\%$ \\ \hline
$h\rightarrow gg$ & $9.1\%$ &-& $14\%$ & $4.1\%$ &-& $3.1\%$ &-\\ \hline
$h\rightarrow c\bar{c}$ & $7.4\%$ &-& $12\%$ & $6.2\%$ &-& $7.6\%$ &-\\ \hline
$h\rightarrow \tau^{+}\tau^{-}$ & $4.2\%$ &-& $5.4\%$ & $14\%$ &-& $3.5\%$ &-\\ \hline
$h\rightarrow ZZ^{\ast}$ & $19\%$ &-& $25\%$ & $8.2\%$ &-& $4.4\%$ &-\\ \hline
$h\rightarrow WW^{\ast}$ & $9.1\%$ &-& $9.2\%$ & $2.6\%$ &-& $3.3\%$ &-\\ \hline
$h\rightarrow \gamma\gamma$ & $34\%$ &-& $34\%$ & $23\%$ &-& $8.5\%$ &-\\ \hline
 \end{tabular}\end {center} \end{table}

In the LRTHM, the modifications of the $hVV (V=Z, W)$ and $hf\bar{f}$ (the SM
fermions pair) couplings can give the extra contributions to the Higgs boson
production processes. On the other hand, the loop-induced couplings, such as
$h\gamma\gamma$ and $hgg$, could also be affected by the presence of top partner,
new heavy charged
gauge bosons and charged scalars running in the corresponding loop diagrams.
Finally, beside the effects already seen in the HS channel due to the exchange
of $s$-channel heavy neutral gauge boson $Z_{H}$, the exchange of top partner
$T$ could also affect the production cross section for the process
$e^{+}e^{-}\to t\bar{t}h$.
All these effects can modify the signal strengths in a way that may be
detectable at the future ILC experiments.

\begin{figure}[tbh]
\begin{center}
\vspace{-1.5cm}
\scalebox{1.2}{\epsfig{file=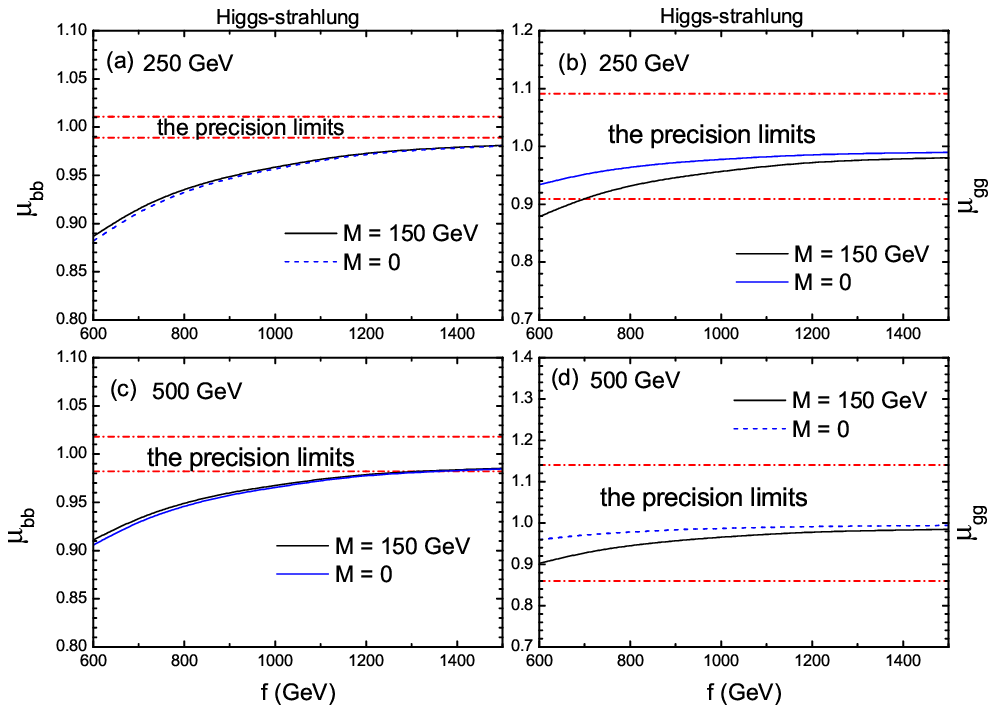}}
\vspace{-2cm}
\scalebox{1.2}{\epsfig{file=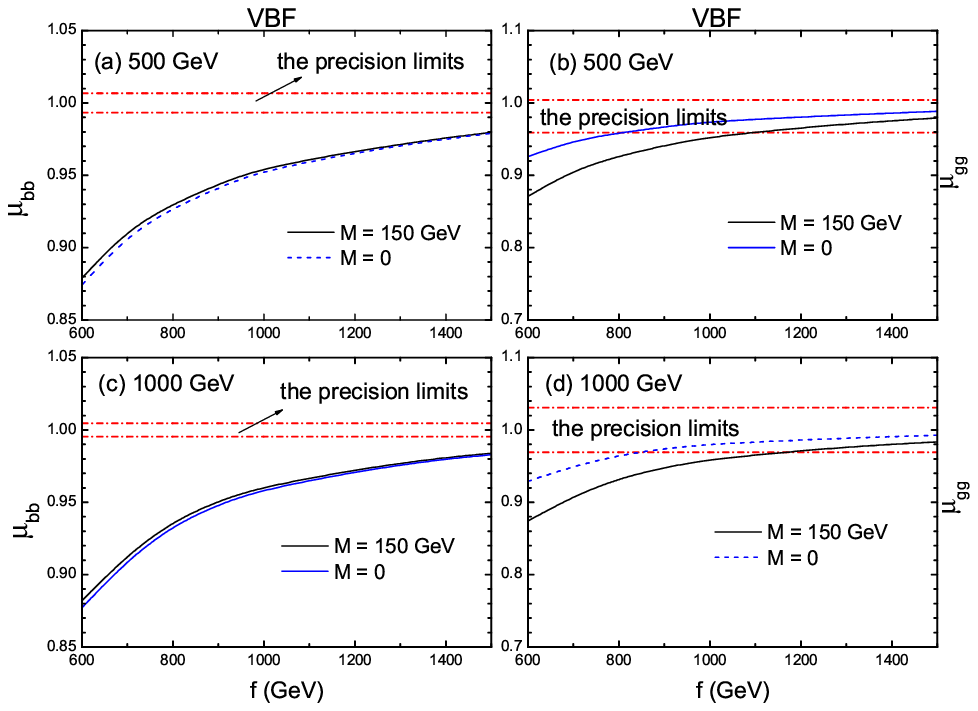}}
\end{center}
\vspace{0.5cm}
\caption{\small Higgs signal strengths $\mu_{i}$ ($i=b\bar{b}, gg$) for
the processes $e^{+}e^{-}\rightarrow Zh$ (upper sub-figures)
and $e^{+}e^{-}\to \nu\bar{\nu}h$
(lower sub-figures) as a function of the model scale $f$ at the ILC with polarized beams.
The dot-dashed red lines represent the experimental precision limits around the SM
expectation according to Table I.}
\end{figure}

In Fig.~6, we show the dependence of the Higgs signal strengths
$\mu_{i}$ ($i=b\bar{b}, gg$) on the parameter $f$ and $M$ for the
HS and VBF processes with polarized beams, where (a) and (c) denote the Higgs
signal strengths $\mu_{b\bar{b}}$ while (b) and (d) denote $\mu_{gg}$.
One can see that
(i) the NP correction becomes smaller rapidly along with the increase of the scale $f$.
(ii) For the HS process, the contributions of the LRTHM
can be detected by the measurement of the $b\bar{b}$ signal rate in the
future ILC experiments. However, it is difficult to observe these effects
via the $gg$ channel due to the relative weak bound.
(iii) For the VBF process, the contribution of the LRTHM can be easily
detected by the measurement of the $b\bar{b}$ signal rate due to the high
expected precision. Meanwhile, this contribution can also be detected by
the measurement of the $gg$ signal rates in most part of the parameter spaces.

In Fig.~7, we show the dependence of the Higgs signal strengths $\mu_{b\bar{b}}$ on the parameter $f$ and $M$
for the process $e^{+}e^{-}\rightarrow t\bar{t}h$.
From Table I we know that the $35\%$ accuracy for top Yukawa couplings expected at $\sqrt{s}=500$ GeV
can be improved to the level of $8.7\%$ at $\sqrt{s}=1$ TeV.
Only for $f=600$ GeV and $M=150$ GeV, the Higgs signal strengths $\mu_{b\bar{b}}$ can reach 0.84 for $\sqrt{s}=500$ GeV.
Thus it is difficult to observe the LRTHM effect on this process at $\sqrt{s}=500$ GeV via the $b\bar{b}$ channel.
For $\sqrt{s}=1$ TeV, however, one can see the magnitude of such correction becomes more sizable for
larger $M$ and lower values of the scale $f$.
For example, for $M=150$ GeV and $f\leq 800$ GeV, the absolute value of $\mu_{b\bar{b}}$ can deviate
from the SM prediction by over $9\%$, which might be detected in the future ILC experiments.

\begin{figure}[tb]
\begin{center}
\scalebox{0.70}{\epsfig{file=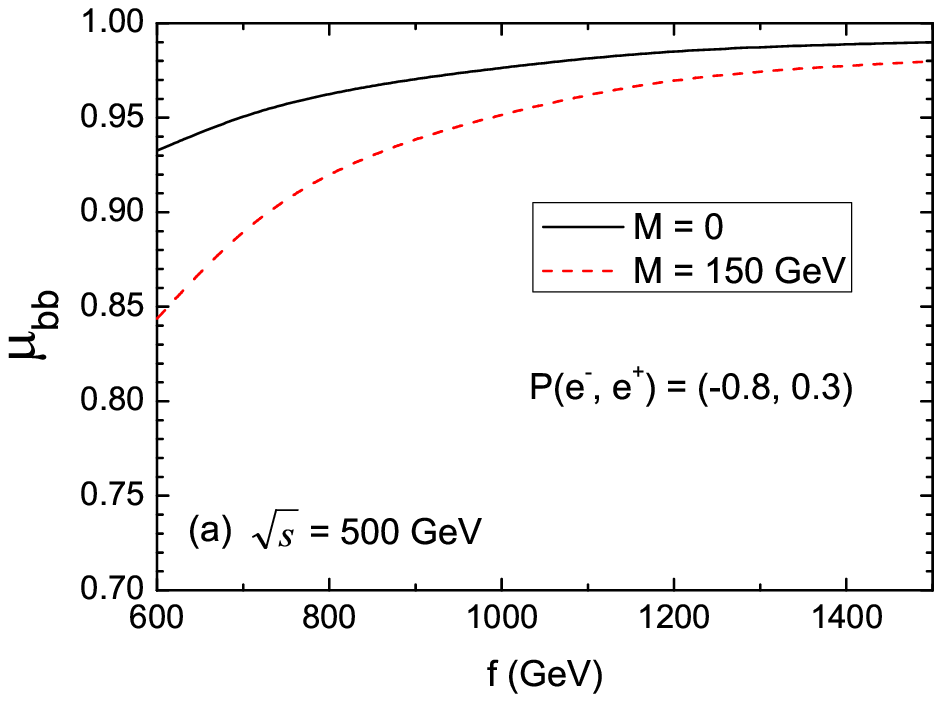} \hspace{-0.5cm}\epsfig{file=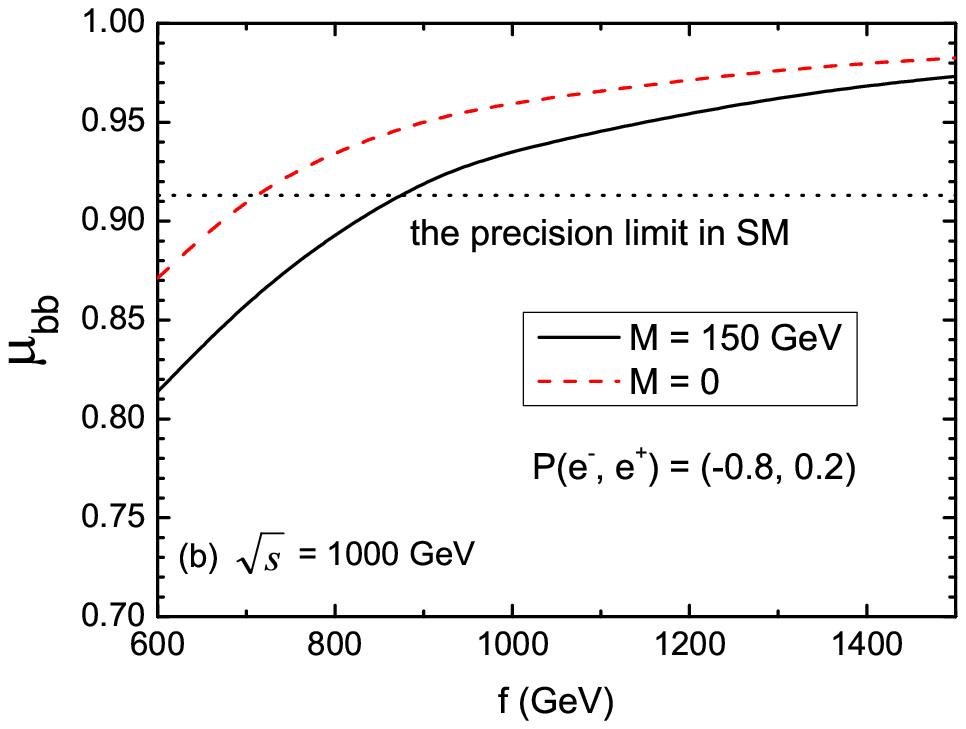}}
\end{center}
\caption{\small Higgs signal strengths $\mu_{b\bar{b}}$ for the process
$e^{+}e^{-}\rightarrow t\bar{t}h$ as a function of the parameter  $f$ and $M$ at
the ILC with polarized beams for $\sqrt{s}=500$ and 1000 GeV.}
\end{figure}

\subsection{Simulated expectations at the ILC}

Now we perform a simulation by using the projected $1\sigma$ sensitivities for the channels
in the HS and VBF processes at the ILC-250 GeV, ILC-500 GeV, and ILC-1000 GeV respectively,
as listed  in Table I \cite{1310.0763,1306.6352}.
The $\chi^{2}$ function can be defined as
\beq
\chi^{2}=\sum_{i}\frac{(\mu_{i}-1)^{2}}{\sigma_{i}^{2}},
\eeq
where $\sigma_{i}$ denotes the $1\sigma$ uncertainty for the signal $i$ in Table 1.
In our evaluations, we take $\chi^{2}-\chi_{min}^{2}\leq$6.18, where $\chi_{min}$ denotes the minimum
of $\chi$ which happens for the largest values of the parameters $f$ and $M$, i.e.
for $M=150$ GeV and $f=1500$ GeV, $\chi_{min}^{2}$=3.33 with the numbers in Table I for the case of ILC-250 GeV,
and $\chi_{min}^{2}$=14.98 with the numbers in Table I for the case of ILC-500 GeV, respectively.
These samples correspond to the $95\%$ confidence level regions in any two
dimensional plane of the model parameters when explaining the Higgs data.

In Fig.~8 we show the allowed region for parameters $M$ and $f$ at the $2\sigma$ level for the cases
of ILC-250 GeV, ILC-500 GeV and ILC-1000 GeV, respectively.
One can see that (i) the constraint has a rather weak dependence on the variation of the parameter $M$;
and (ii) the allowed region of the scale parameter $f$ for the case of the  ILC-1000 GeV become much narrow
than that for ILC-250 GeV.
At the $2\sigma$ level, for example, the value of $f$ must be larger than 1400 GeV for ILC-1000 GeV, while
the lower limit is 1150 GeV for the cas of ILC-250 GeV.
The above bounds from the proposed ILC measurements may be much stronger than that for the LHC Higgs data \cite{liu-1311}.

\begin{figure}[ht]
\begin{center}
\vspace{0.5cm}
\centerline{\epsfxsize=7cm\epsffile{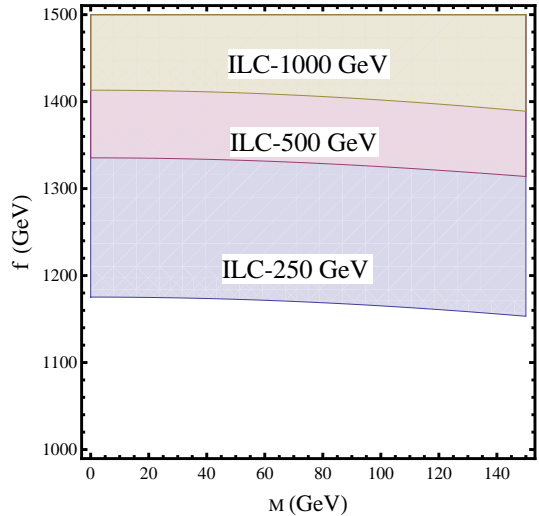}}
\caption{The allowed region for parameters $f$ and $M$ are shown at the $2\sigma$ level for the cases of
ILC-250 GeV, ILC-500 GeV and ILC-1000 GeV, respectively. }
\label{fig:fig8}
\end{center}
\end{figure}

\section{ Conclusions}

The LRTHM is a concrete realization of the twin Higgs mechanism, which provides an
alternative solution to the little hierarchy problem.
In this work, we studied three Higgs boson production processes $e^{+}e^{-}\rightarrow Zh$,
$e^{+}e^{-}\rightarrow \nu\bar{\nu}h$ and $e^{+}e^{-}\rightarrow t\bar{t}h$ in the framework of the LRTHM.
We calculated the production cross sections for three processes with and without the polarized beams, the
relative corrections with the polarized beams for three energy stages.
We also studied the signal rates with the SM-like Higgs boson decaying to
$b\bar{b}$ and $gg$, and performed a simulation by using the projected $1\sigma$ sensitivities
as listed  in Table I. Our numerical results show that:
\begin{enumerate}
\item
For the considered three processes, the production cross sections with polarized beams are larger
than those with unpolarized beams, which are more sensitive to the LRTHM;

\item
In a large part of the allowed parameter space, the LRTHM can generate
moderate contributions to the HS and VBF processes.
For the process $e^{+}e^{-}\rightarrow t\bar{t}h$, we found that in certain
regions of parameter space (for larger $M$ and lower
values of the scale $f$), the absolute value of the Higgs signal strength
$\mu_{b\bar{b}}$ can deviate
from the SM prediction by over $8.7\%$, and thus may be detectable at the
future ILC for $\sqrt{s}=1$ TeV with polarized beams $P(e^{-},e^{+})=(-0.8,0.2)$.

\item
The future ILC experiments can give strong limit on the scale parameter $f$. For
the case of ILC-250 (1000) GeV, the value of $f$ must be larger than 1150 (1400)
GeV at the $2\sigma$ level.

\end{enumerate}

\begin{acknowledgments}
We would like thank Shufang Su for providing the CalcHep Model Code.
 This work is supported by the National Natural Science
Foundation of China under the Grant No. 11235005, the Joint Funds
of the National Natural Science Foundation of China (U1304112), and by
the Project on Graduate Students Education
and Innovation of Jiangsu Province under Grant No. KYZZ-0210.

\end{acknowledgments}


\end{document}